# Interlayer Tunneling of Quasiparticles and Cooper Pairs in Bi-2212 Single Crystal Whiskers

Yu. I. Latyshev[a,b], V. N. Pavlenko[a,b], S.-J. Kim[b], T. Yamashita[b], L. N. Bulaevskii[c], M. J. Graf[c], A. V. Balatsky[c], N. Morozov[c], and M. P. Maley[c]

[a]Institute of Radio-Engineering and Electronics Russian Acad. of Sciences, 11 Mokhovaya str., 103907 Moscow, Russia

[b]Research Institute of Electrical Communication, Tohoku University, 2-1-1, Katahira, Aoba-ku, Sendai 980-8577, Japan

[c]Los Alamos National Laboratory, Los Alamos, New Mexico 87545, USA

The interlayer tunneling has been studied on high quality Bi-2212 stacks of micron to the submicron lateral size. We found that low temperature and low voltage tunneling *I-V* characteristics can be self-consistently described by Fermi-liquid model for a *d*-wave superconductor with a significant contribution from coherent interlayer tunneling. The gap and pseudogap interplay with variation of temperature and magnetic field has been extracted from the *I-V* characteristics. We consider also the role of charging effects for submicron stacks.

## 1. INTRODUCTION

The interlayer tunneling of both quasiparticles and Cooper pairs may be studied in layered high-$T_c$ cuprates by measuring the *I-V* characteristic of the *c*-axis current [1]. We report here our recent measurements of interlayer tunneling characteristics in small Bi-2212 stacks with emphasis on significant contribution of coherent processes to the interlayer tunneling and on *d*-wave symmetry of the order parameter.

## 2. RESULTS AND DISCUSSION

As a base material for junction fabrication we used Bi-2212 single crystal whiskers grown by impurity free method [2]. Recently they have been characterized [2] as a very perfect crystalline object. For the fabrication of stacked junction we used the conventional FIB machine of Seiko Instruments Corp., SMI 9800 (SP) with $Ga^+$-ion beam. The junctions have been fabricated by double-sided processing of whisker with FIB. The details of fabrication steps are described in [3]. Parameters of the stacks are listed in the Table 1.

### 2.1. Low temperature and voltage *I-V* characteristics.

Figures 1a and 1b show the *I-V* characteristics of samples #3 and #4. The fully superconducting overlap geometry of the stack was used to suppress the effects of quasiparticle injection on the tunneling characteristics, usually occurring in junctions of the mesa type with a normal metal top electrode [1,4]. We also substantially reduced the effects of selfheating in our submicron mesa junctions. Self-heating manifests itself in the form of an S-shaped $I$-$V$ curve near the gap voltage $V_g$ [5]. The measured temperature dependence of the $c$-axis resistivity of the stack was typical for slightly overdoped Bi-2212 crystals with $T_c \approx 77$ K [6]. The critical current $I_c$ was determined from the $I$-$V$ characteristics as the current of switching from the superconducting to the resistive state, averaged over the stack. The $c$-axis critical current density, $J_c$, for the junctions with in-plane area $S > 2$ μm$^2$ was typically 1 kA/cm$^2$ at T = 4.2 K [5]. The superconducting gap (pseudogap) voltage of the stack, $V_g$, was determined from the $I$-$V$ characteristics as the voltage at the maximum of the $dI/dV$. The gap of the intrinsic junction, $2\Delta_0 \approx eV_g / N$, where $N$ is the number of elementary layers in the stack, reaches value as high as 50 meV (see also [7]). The multibranched structure, which is clearly seen in Fig. 1a corresponds to subsequent transition of the intrinsic junctions into the resistive state for increasing voltage [8]. At voltages $V > V_g$ all junctions are resistive. In downsweep of voltage, starting from $V > V_g$ the $I$-$V$ curve is observed in the all junction resistive state. Corresponding quasiparticle conductivity, $\sigma_q$, thus can be defined directly from that part of the $I$-$V$ curve. The Ohmic resistance, $R_n$, at $V > V_g$ is well defined (Fig. 1b).

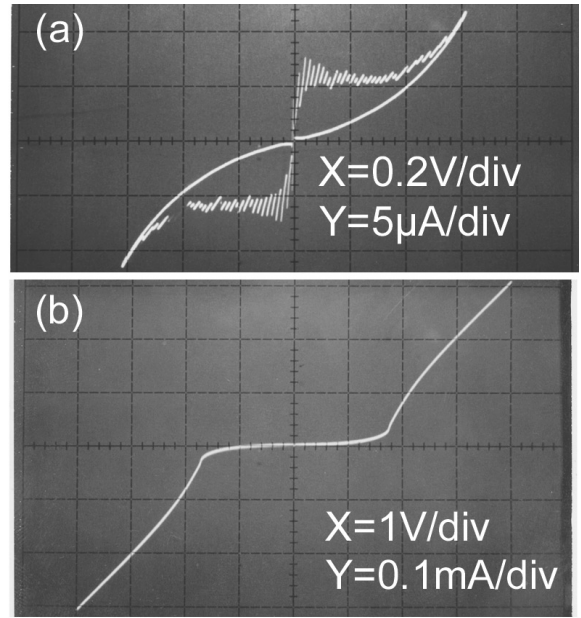

Figure 1. *I-V* characteristics of the Bi-2212 stacks in (a) enlarged scale for sample #3 and (b) extended scale for sample #4. *T*=4.2 K.

Table 1
Parameters of Bi-2212 stacked junctions.

| No | S (μm$^2$) | N | $V_g$ (V) | $I_c$ (μA) | Notes |
|---|---|---|---|---|---|
| 1 | 0.4 | 70 | 2.4 | 0.1 | |
| 2 | 2.0 | 65 | 1.3 | 12 | |
| 3 | 1.5 | 50 | 1.1 | 6 | |
| 4 | 0.6 | 35 | 1.7 | 0.25 | |
| 5 | 0.3 | 50 | 2.2 | 0.07 | |
| 6 | 36x0.5 | 50 | 3.0 | 14 | Array 6x6 junctions |
| 7 | 400 | 40 | 0.8 | 1000-2000 | Data taken from Ref. [1] |
| 8 | 0.4 | 70 | 2.4 | ~0.05 | |

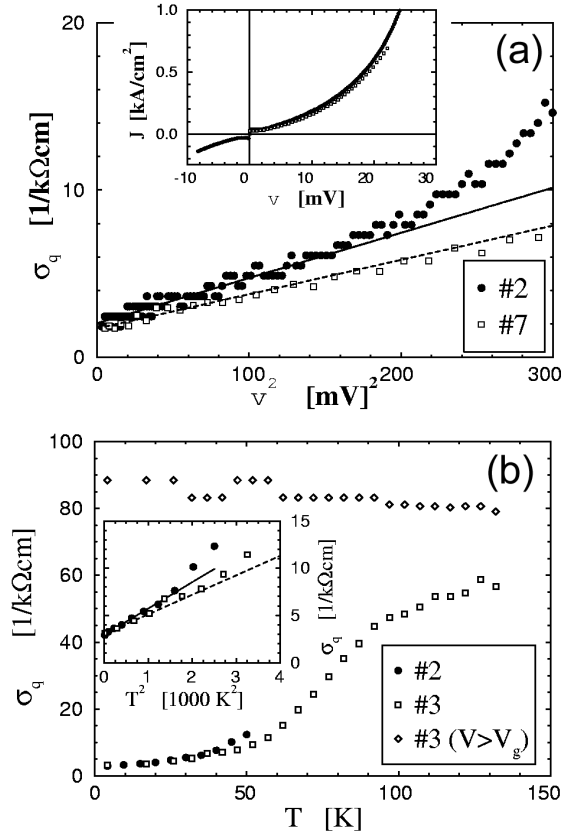

Figure 2. Quasiparticle dynamic conductivity (a) vs $v^2 = V^2/N^2$ at $T$=4.2 K and (b) vs $T$ for $v > V_g/N$ and $v \to 0$. Inset in (a): corresponding $J$-$v$ curves; (b): $\sigma_q$ vs $T^2$ at $v \to 0$. Lines are fits for $v < 10$ mV and $T^2 < 1000$ K$^2$, correspondingly.

This resistance is nearly temperature independent (Fig. 2b) and corresponds to the conductivity $\sigma_n(V > V_g) \approx 80$ (kΩ cm)$^{-1}$ for energies above superconducting gap and pseudogap.

We found out that interlayer tunneling $I$-$V$ characteristics at low temperatures essentially differ from those of conventional Josephson junctions between $s$-wave superconductors. We specify [9]: (1) strong disagreement with Ambegaokar-Baratoff (A-B) relation, $J_c^{AB}(0) = \pi \sigma_n \Delta_0 / 2 e s$, with $s$ the spacing between intrinsic superconducting layers (15.6 Å); (2) quadratic dependences of quasiparticle conductivity $\sigma_q(V, T)$ on $V$ and $T$ (Fig. 2): $\sigma_q(V, 0) = \sigma_q(0, 0) ( 1 + \alpha V^2)$, $\sigma_q(0, T) = \sigma_q(0, 0) (1 + \beta T^2)$ with $\alpha = 0.014 \pm 0.003$ (meV)$^{-2}$, $\beta = (6 \pm 2) \times 10^{-4}$ K$^{-2}$, (3) nonzero and universal value of $\sigma_q(0, 0) \approx 2$ (k$\Omega$ cm)$^{-1}$. The last two features have been recently reproduced in Ref. [10].

We found empirically the modified relation of the A-B type,

$$J_c(0) \approx \pi \sigma_q(0, 0) \Delta_0 / 2 e s \quad (1)$$

and scaling relation between $\alpha$ and $\beta$, $\beta / \alpha = $ const $\approx 4 \pm 2$. It was shown that all these features can be described self-consistently by Fermi-liquid model for quasiparticles in clean $d$-wave superconductor with resonant scattering [9]. Impurity scattering leads to the formation of gapless state near the node directions $\varphi_g$ at angles $\varphi_g \pm \varphi_0/2$, with $\varphi_0 \sim \gamma / \Delta_0$ where $\gamma$ is the impurity bandwidth of quasiparticles. That results in a nonzero density of states at zero energy, and leads to a universal quasiparticle interlayer conductivity $\sigma_q(0, 0)$. It was shown also [9] that the values of $\sigma_q(0, 0)$, $J_c(0)$ and coefficients $\alpha$ and $\beta$ are strongly dependent on the coherency of interlayer tunneling. If to denote the weight for in-plane momentum conserving (coherent) tunneling as $a$ and for incoherent as $(1 - a)$, the ratio $J_c(0) / \sigma_q(0, 0)$ is expressed as follows [9]

$$\frac{J_c(0)}{\sigma_q(0,0)} \approx \frac{\pi \Delta_0}{2es} \frac{a + (1-a) \Delta_0 / \varepsilon_F}{a + (1-a) \gamma / \varepsilon_F} \quad (2)$$

with $\varepsilon_F$ the Fermi energy. Correspondingly for coefficients $\alpha$ and $\beta$ it was obtained [9]

$$\alpha \approx 1/8 \gamma^2 [1 + (1/a - 1) \gamma / \varepsilon_F]$$
$$\beta \approx \pi^2 / (18 \gamma^2) [1 + (1/a) - 1) \gamma / \varepsilon_F] \quad (3)$$

One can see that Eq (2) turns to the experimentally found relation (1) only for significant contribution of coherent tunneling, $a \gg \max \{\Delta_0 / \varepsilon_F, \gamma / \varepsilon_F\}$. From experimental value for $\beta$ we can estimate $\gamma$ from Eq. (3) to be $\gamma \approx 3$ mV. Then for $\Delta_0 / \varepsilon_F \approx 0.1$ we can get estimation for $a$, $a \gg 0.1$. Eqs. (3) give $\beta / \alpha = 4 \pi^2 / 9$ in a reasonable agreement with experiment.

**2.2. Charging effects.**

As is well known, the Josephson tunneling can be suppressed by the Coulomb blockade effect in small junctions [11], when the charging energy $E_c$ becomes comparable with the Josephson coupling energy $E_J$.

For the submicron junctions we clearly observed well-defined tunneling charcteristics with superconducting gap (Fig. 1b). We found also a number of new features [5], typical for the single Cooper-pair tunneling in small tunnel junctions [12]. The I-V characteristics have no hysteresis and multibranched structure, the critical current has a finite slope increasing with a temperature, the periodic structure of current peaks develops on the I-V curves at low temperatures (Fig. 3), $J_c(0)$ is reduced.

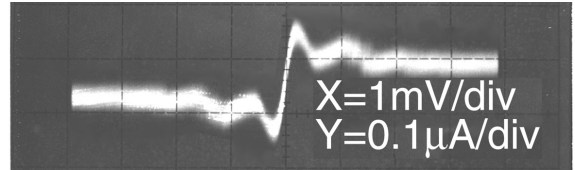

Figure 3. Extended scale $I$-$V$ characteristics of submicron Bi-2212 stack #5. The period of structure $\Delta V$ corresponds to the charging energy for single Cooper pair transfer through the stack.

The scale of charging energy observed is consistent with charge soliton model [13] considered for our stacks [5].

**2.3. Gap and pseudogap spectroscopy.**

The interlayer tunneling $I$-$V$ characteristics in small stacked junctions provide an important possibility of gap and pseudogap spectroscopy in layered high-$T_c$ cuprates. In comparison with other widely used methods for studies gap and pseudogap like ARPES or STM (see as a review Ref. [14]) which are in fact surface methods, the interlayer tunneling spectroscopy gets an information from the body of the single crystal. Our submicron Bi-2212 stacks show high quality tunneling characteristics. Both the gap and the pseudogap are clearly seen in the *I-V* characteristics. Fig. 4a shows temperature dependence of gap and pseudogap. Starting from low temperatures a gap goes down more rapidly than usual BCS dependence (solid line). We defined $T_c$ at a point (77 K) where $I_c(T)$ turns to zero. At $T > 65$ K the gap evolves into the pseudogap that was observed up to 160 K. The maximum pseudogap value $2\Delta_p$ at ~ 110 K is about 20% higher than superconducting gap $2\Delta_0$. We have never observed simultaneously a gap and a pseudogap at the interval 65 K < $T$ < 77 K. The most interesting observed feature is that the pseudogap abruptly goes down approaching $T_c$ from high temperatures. It may be an indication of the anti-coexistence of the gap and the pseudogap. That type of temperature behaviour has been reproduced recently on break Bi-2212 junctions [15], but is in conflict with STM measurements on Bi-2212 [16], where neither gap nor pseudogap variation with temperature has been reported. To clarify the point we undertaken studies of the gap and pseudogap dependences on magnetic field *H // c* up to 6T (see Fig. 4b). We found that the gap is suppressed by magnetic field. To the contrast the pseudogap enhances by H. That observation seems to exclude pseudogap origin due to superconducting fluctuations [17] or due to existence of preformed Cooper-pairs [18] above $T_c$ and points out to the different origin of ordering for the gap and for the pseudogap formation.

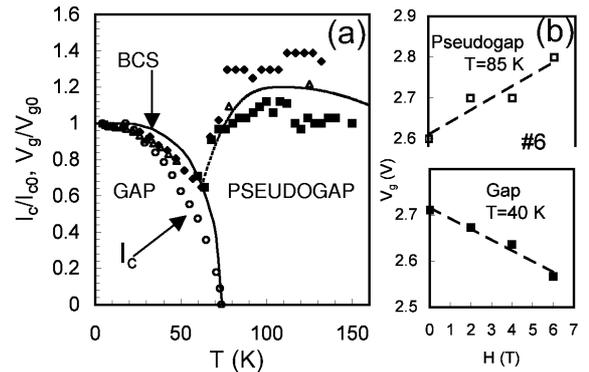

Figure 4. Gap and pseudogap dependences on $T$ (a) and magnetic field $H//c$ (b). Denotes of the panel (a) correspond to the following samples: ♦ - #3, Δ - #8, ■ - #1. Line for pseudogap temperature dependence is guide to eye.

## 3. CONCLUSIONS

We studied interlayer tunneling on small stacked junctions of different size. The results obtained point out to the *d*-wave symmetry of order parameter, to the significant contribution from coherent interlayer tunneling, to the "non-superconducting" origin of the pseudogap. We uncovered also the influence of charging effects on interlayer tunneling.

**Acknowledgments.**

We thank A. M. Nikitina for providing us with Bi-2212 single crystal whiskers. This work was supported by CREST, the Japan Science and Technology Corporation, and the Russian State Program on HTS under grant No. 99016.